\documentclass[final,5p,twocolumn]{elsarticle}

\usepackage[cp1251]{inputenc}
\usepackage[english]{babel}

\usepackage{graphicx}
\usepackage{epstopdf}

\usepackage{amssymb}
\usepackage{amsmath}

\journal{Nuclear Instruments and Methods in Physics Research A}

\begin{document}

\renewcommand{\figurename}{Fig.}

\begin{frontmatter}



\title{A simplified spectrometer based on a fast digital oscilloscope for the
measurement of high energy $\gamma$-rays}


\author[mipt]{S.\,S.\,Markochev\corref{cor1}}
\ead{sergey.markochev@gmail.com}

\author[msu]{N.\,V.\,Eremin}

\address[mipt]{Moscow Institute of Physics and Technology, Dolgoprudny 141700,
Russian Federation}

\address[msu]{Skobeltsin Institute of Nuclear Physics, Lomonosov Moscow State
University, Moscow 119991, Russian Federation }

\cortext[cor1]{Corresponding author}

\begin{abstract}
A simplified digital spectrometer for the study of $\gamma$-rays with
energies up to $\sim100$ MeV is presented and tested. The spectrometer is only consisted of a
fast digital oscilloscope and three scintillation detectors which can work
in single or in coincidence modes: two BGO-detectors comprising
$\varnothing\,7.62\times7.62$~cm BGO-crystalls and one
plastic detector which includes an organic polystyrene-based scintillator.
The basic properties of the spectrometer (energy resolution, time resolution,
$\gamma$-rays detection efficiency) were studied exhaustively also
using a Geant4-based Monte-Carlo simulation. 
Several numerical algorithms for processing of waveforms in offline mode were
proposed and tested to perform digital timing, pulse area measurement and
processing of pile-up events without rejection. As a result, the spectrometer
demonstrated $\sim10\%$ better energy resolution than was obtained by
a common 10-bit CAMAC ADC with the same detectors. 
And the developed algorithm based on the pulse shape analysis for
processing of pile-up events showed high efficiency under severe
conditions (the portion of pile-ups contained $\sim30\%$).
The measured maximum counting rate of the spectrometer was 
$1.8\times10^5$~waveforms/sec.

\end{abstract}


\end{frontmatter}


\section{Introduction}

The study of energy and angular spectra of high energy $\gamma$-rays
emitted in the spontaneous fission of heavy nuclei remains one 
of the most attractive ways to create a complete picture of the evolution of the
fissioning system (emission of prescission $\gamma$-rays \cite{Glassel_1989},
excitation of giant dipole resonances in fission fragments
\cite{Ploeg_1995, Glassel_1989}), and to search for exotic
radioactivity of heavy nuclei \cite{Otsu_1992, Giorgini_2009}.
While the emission process of $\gamma$-rays and neutrons with energies
$E_{\gamma}\leq10\ldots20$~MeV has been studied well enough, the nature
of $\gamma$-rays with energies $E_{\gamma}>20$ MeV accompanying the
spontaneous fission of heavy nuclei should be clarified.
Actually, Kasagi et al.\ \cite{Kasagi_1989} reported the emission of 
high energy $\gamma$-rays with energies up to 160~MeV from the spontaneous
fission of $\rm ^{252}Cf$. Similar results were obtained by Pandit et
al.\ \cite{Pandit_2010} up to 100~MeV and by Ploeg et al.\
\cite{Ploeg_1992} up to 40~MeV. The measured value of $\gamma$-ray emission probability  
at 40~MeV was $\rm\sim10^{-7}\,photon/(MeV\cdot fission)$ and at 160~MeV
was $\rm\sim10^{-8}\,photon/(MeV\cdot fission)$. These results are in
contradiction with those were obtained by other three research groups:
Pokotilovsky \cite{Pokotilovsky_1990} established upper limits of
$6\times10^{-9}$ and $\rm1\times10^{-9}\,photon/(MeV\cdot fission)$ at
$E_{\gamma}=40$ and 100~MeV respectively, and Luke et al.\
\cite{Luke_1991} and Varlachev et al.\ \cite{Varlachev_2005} established upper
limits of $1.8\times10^{-6}$ and $\rm1.2\times10^{-10}$ photon
per fission for the integrated yield of $\gamma$-rays with energies
$E_{\gamma}>30$~MeV and $E_{\gamma}>38$~MeV respectively.

In these experiments large complex setups were used. For example, in the
paper of Kasagi et al.\ \cite{Kasagi_1989} high energy $\gamma$-rays were
detected by a seven-element $\rm BaF_2$ array (each element had a size of $\rm
37\,cm^2\times 20\,cm$), which was controlled by analog CAMAC electronics. In the
other experiments efforts inclined mostly to construct setups using large
volume $\rm NaI$ and $\rm BaF_2$ detectors with an active anticoincidence
shield \cite{Pokotilovsky_1990, Luke_1991, Varlachev_2005} or using
multidetector arrays \cite{Pandit_2010, Ploeg_1992}, but application of analog electronics for data acquisition and data
processing continued.  
To analyse the reasons why the experimental groups
\cite{Kasagi_1989, Pokotilovsky_1990, Luke_1991, Ploeg_1992, Varlachev_2005,
Pandit_2010} obtained so different results, we note here the crusial problems
of the experimental study of this phenomenon: such rare nuclear events occur on
the probability level of $\rm\sim10^{-7}\,photon/(MeV\cdot fission)$,
and results are strongly influenced by cosmic background and the pile-up effect.
We suggest the analog apparatus has to be changed by the digital one to move on.
Rapid develoment of fast digital oscilloscopes and
appearance of new scintillation materials can really improve and simplify the
experimental setup.
All modules of the analog electronics can be replaced by only one fast digital
storage oscilloscope and BGO-scintillators with smaller sizes can be used
instead of large volume $\rm NaI$ and $\rm BaF_2$ crystals.

In this paper the detailed description of the $\gamma$-ray spectrometer based on
a fast digital oscilloscope is presented together with numerical
algorithms for timing, pulse area measurement and processing of pile-up events
without rejection. 
This spectrometer was used to study high energy $\gamma$-rays emission
accompanying the spontaneous fission of heavy nuclei \cite{IJMP_2010, JOP_2011}. Application
of a fast digital oscilloscope simplifies the setup greatly and achives much better performance
over the analog apparatus in the case of long time experiments. In addition, new
numerical algorithm for processing of pile-up events without rejection was successfully
tested, and calibration of the spectrometer up to $\sim100$~MeV with results of
the time-of-flight method implementation to distinguish $\gamma$-rays and
neutrons are presented.

\section{Setup and software}
\label{setup_description}

\begin{figure}
\centering
\includegraphics[width=0.48\textwidth]{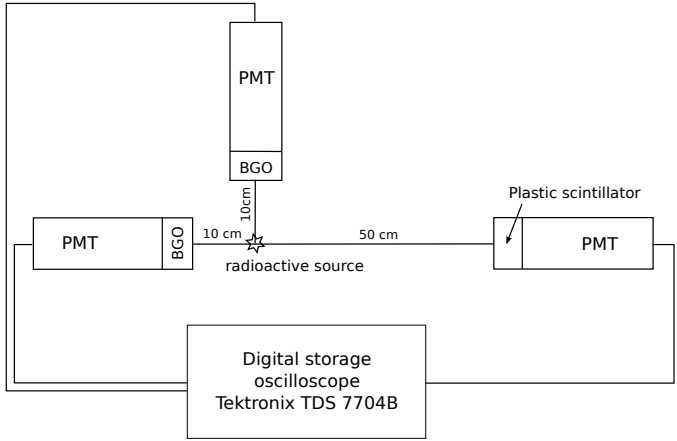}
\caption{Schematic diagram of the spectrometer for the measurement of high
energy $\gamma$-rays. Digital oscilloscope: Tektronix TDS 7704B (4~inputs, 7
GHz bandwidth, 20 GS/s max sampling rate). PMT: Photonis
XP4312 photomultiplier tube (risetime of 2.1 ns).}
\label{figure_experimental_setup}
\end{figure}

\figurename\ref{figure_experimental_setup} shows a schematic diagram of the
spectrometer, which consists of two BGO-detectors, one 
plastic detector and a fast digital storage oscilloscope Tektronix TDS 7704B (4
channel inputs, 7 GHz bandwidth, 20 GS/s maximum sampling rate, 8-bit ADC).
The BGO-detectors were composed of $\varnothing\,7.62\times7.62$~cm BGO crystals
coupled with Photonis XP 4312 photomultiplier tubes (PMT). 
The plastic detector was composed of a polystyrene-based scintillator
with a size of $\varnothing\,6\times2$ cm and Photonis XP 4312 PMT. 
The detectors were attached to rails, which
determined angles between the BGO-detectors and the plastic detector. The
values of the angles were equal to $\rm 180^o$ and $\rm 90^o$
(\figurename\ref{figure_experimental_setup}). A radioactive source ($\rm
^{252}Cf$, PuBe and so on) was placed at a distance
of 10~cm from the BGO-detectors and at 50~cm from the plastic detector.

Anode pulses of the PMTs were sent directly to the inputs of
the digital oscilloscope, where continuous output signals coming from the
detectors were digitized into discrete waveforms and stored in the
oscilloscope memory after triggering condition was satisfied.
Later the waveforms were saved to the oscilloscope's hard disk and were
transferred to a PC through a LAN interface, where they were processed by
numerical algorithms in offline mode. Time duration of the waveforms was set
to 2~$\mu$s, which was enough to analyse the shape of BGO-pulses (the scintillation
primary decay time of a BGO crystal is $\approx300$~ns). The sampling time
interval was chosen to 0.2~ns ($10^4$ sampling points per waveform).

All operations of the digital oscilloscope (waveforms acquisition, data
acquisition, oscilloscope parameters setup and others) were controlled by special
software written in C++ using a Plug$\&$Play oscilloscope driver. 
The application connected to the oscilloscope by a TekVisa interface and
could perform the following operations:

\begin{itemize}
  \item to collect and to save pulses waveforms to a hard disk;
  \item to save trigger time for each detected pulse (this is needed for
  time series analysis);
  \item to measure a counting rate on each channel;
  \item to reset parameters of the oscilloscope and to save they to a file;
  \item to input, to save and to load an order of operations to perform.
\end{itemize}

Before measurements the oscilloscope was set up to required states 
in which the oscilloscope will perform operations in the future. These states of
the oscilloscope were saved to files. Later a sequence of operations was
inputted in the application, so that the oscilloscope could load the desired
states in time. After the application was started the spectrometer was needed
nobody to control. This method allowed us to conduct experiments without any
human control for up to 4~months with periodic pauses for calibration of the spectrometer.

High energy $\gamma$-rays could be detected in two ways: in single mode or in
coincidences with low energy $\gamma$-rays and neutrons, which were detected
by the plastic detector. Low energy $\gamma$-rays and neutrons were clearly
distinguished by the time-of-flight histogram constructed from the time-mark
differences between the input waveforms from the plastic detector and any of
the BGO-detectors.

\subsection{Timing analysis algorithm}
\label{gauss_smooth}

\begin{figure}
\centering
\includegraphics[width=0.48\textwidth]{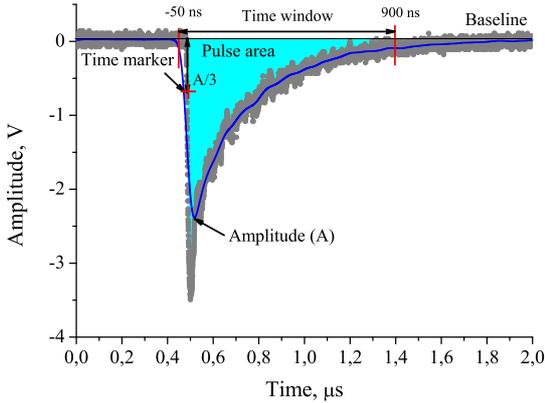}
\caption{The scheme of a BGO-pulse processing. Gray dots denote original
points of the pulse; blue line --- the same pulse smoothed by a Gaussian
function with a parameter $\sigma_1=30$~ns; the pulse area is filled
with light blue color; red ticks denote the boundaries of the integration time
window; red cross --- the time-mark.}
\label{figure_pulse_processing}
\end{figure}

In this paper the aim of timing analysis was to acquire
the ``arrival times'' of detected pulses, which were used for
pulse area measurements and to distinguish $\gamma$-rays from
neutrons by the time of flight. In Ref.~\cite{Nissila_2005} it was demonstrated
that the constant fraction timing method is the best one, which achieves the highest time
resolution. According to this method the time-mark is placed
on the leading edge of a pulse at the certain constant fraction level of
its full amplitude. The level of $1/3$ was chosen in this work
(see \figurename\ref{figure_pulse_processing}). 

Timing analysis included smoothing of the pulses and search for
places for the time-marks on the fronts of the smoothed pulses (see
\figurename\ref{figure_pulse_processing}). 
The smoothing was performed by a convolution of the pulses with a Gaussian function.
A parameter $\sigma_1=30$~ns was chosen for the amplitude measurements 
and $\sigma_2=2$~ns for the time-mark place searches (a
Gaussian function with the parameter value $\sigma_1$ was a good approximation
of the front of a BGO-detector pulse, whereas smoothing of the pulses using the parameter value
$\sigma_2$ showed the best time resolution). A source of $\rm ^{252}Cf$ with an activity of
$3.6\times10^5$~fission/sec was used to test the algorithm. 
The time-difference spectrum of $\gamma$-$\gamma$ and neutron-$\gamma$
coincidences from the $\rm ^{252}Cf$ source is presented in
\figurename\ref{figure_spectrum_coincidences}. 
The time resolution of the spectrometer was 2.2~ns (the risetime of PMT
was 2.1~ns). Is was measured as full-width at half-maximum (FWHM) of the
$\gamma$-$\gamma$ peak, which was approximated by a Gaussian function. 
The bump on the left side of the $\gamma$-$\gamma$ peak corresponds to
events when the BGO-detector registered a neutron and
the plastic detector registered a $\gamma$-quantum.

\subsection{Pulse area measurement algorithms}

The simplest method to measure $\gamma$-ray energy is to calculate
area under a pulse within a fix integration time window. It is convenient to
dispose this time window relative to the time-mark of a pulse. 
The boundaries of the integration time window was chosen so that the
BGO-detectors achieved the best energy resolution for standard
$\gamma$-ray sources: $\rm ^{137}Cs$, $\rm ^{60}Co$ and PuBe. The best value of
the energy resolution was obtained using the time window $-50\ldots900$~ns relative to the time-mark, which
corresponds to the ``0~ns'' mark.
For the plastic detector the time window $-8\ldots30$~ns was chosen
relative to the time-mark to capture a pulse completely.

Another approach to pulse area measurement is to use the model description of
the pulse shape of inorganic scintillators consisting of the sum of three exponents: the
first exponent describes the front of a pulse, the other two --- ``fast''
and ``slow'' components of scintillation light emission, which describe the tail
of a pulse:

\begin{figure}
\centering
\includegraphics[width=0.48\textwidth]{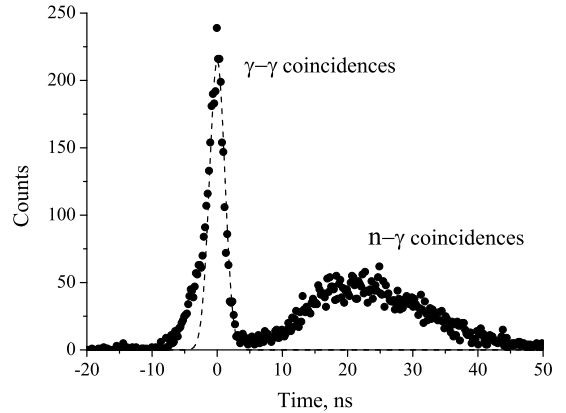}
\caption{The experimental time-difference spectrum from a $\rm ^{252}Cf$ source
obtained in coincidences between one BGO- and the plastic detectors. The
$\gamma$-$\gamma$ peak was approximated by a Gaussian function to extract the time resolution of
the spectrometer (FWHM = 2.2~ns).}
\label{figure_spectrum_coincidences}
\end{figure}

\begin{equation}
s(t) = A_1\cdot e^{-\frac{t}{\tau_1}} + A_2\cdot
e^{-\frac{t}{\tau_2}} - (A1 + A2)\cdot
e^{-\frac{t}{\tau_{front}}}
\label{approximation_curve}
\end{equation}

\noindent where $A_1$ and $A_2$ --- the relative magnitudes of the fast and
slow components; $\tau_{front}$ --- the pulse
risetime; $\tau_1$ and $\tau_2$ --- the fast and slow decay time constants.
The values of $A_i$ and $\tau_i$ can be found by approximation of the
pulse shapes.
Hence, a pulse area equals to the integral of function from
Eq.~\eqref{approximation_curve} from 0 to $+\infty$:

\begin{equation}
E(t) = A_1\cdot(\tau_1 - \tau_{front}) + A_2\cdot(\tau_2 - \tau_{front}) 
\end{equation}

This method requires much more calculation time than the first one and
shows worse energy resolution on standard $\gamma$-ray sources. Nevertheless,
the model of the shape of a BGO pulse is useful if pile-up events occur. In this
case there is a high probability that more than one pulse is found within the
integration time window, so the pulse area measurement cannot be done.
To narrow the integration time window pulses were smoothed as
it was described in paragraph \ref{gauss_smooth} (the parameter value $\sigma_1$
was used).
The energy resolution of the BGO-detectors on the 4.43~MeV $\gamma$-line from
a PuBe source was $\approx8\%$ using the integration time window $0\ldots300$~ns
relative to the time-mark and $\approx9\%$ using the $0\ldots200$~ns one.

\subsection{Pile-ups processing algorithm}
\label{algorithm_of_processing_of_pileups}

The study of high energy $\gamma$-rays emission accompanying the
spontaneous fission of heavy nuclei requires processing of pile-up events.
There are two reasons for this: i) it is important to prove that pulses corresponding
to high energy $\gamma$-rays are not overlapped by low energy $\gamma$-quanta;
ii) rejection of pile-up events reduces the measured probability
of $\gamma$-rays emission.
So, the use of fast signal digitization coupled with numerical processing of
waveforms can solve this problem.  

An example of a numerical algorithm for processing of pile-up events was
described in Ref.~\cite{Belli_2008}. The pulses were fitted by means of the
Levenberg–Marquardt method with six free parameters to reconstruct the shapes of
pulses overlapped in pile-ups. Once the first pulse was reconstructed using 
Eq.~\eqref{approximation_curve}, it was subtracted from the original pulse and
so on. The shape of scintillation pulses in Ref.~\cite{Belli_2008} depended
on whether a neutron or a $\gamma$-quantum was detected. In this paper we
present a new algorithm for processing of pile-ups. As BGO scintillator has 
one primary decay time constant $\tau=300$~ns ($A_{60\,ns}/A_{300\,ns}\sim0.1$
from Ref.~\cite{Avdeichikov_1994}) and considering only the pulse tail, 
Eq.~\eqref{approximation_curve} can be reduced to a very short form with only
one free parameter $A$:

\begin{equation}
E(t) = A\cdot e^{-\frac{t}{\tau}}
\label{simple_curve}
\end{equation}

\noindent where $A$ --- the pulse amplitude; $\tau$ is the primary decay
time constant.
Hence, the non-linear curve-fitting problem originated from
Eq.~\eqref{approximation_curve} does not need to be solved
what gives greater computational speed of the algorithm. The algorithm puts
the time-marks and measures the pulse area if the time difference between
the arrival times for every next two pulses in a waveform exceeds 300~ns (to
measure the pulse area) and the contribution to the current pulse area from the
previous pulses is less than 20\% (to apply the algorithm of timing analysis).
The algorithm consisted of the following steps:
\begin{itemize}
  \item The waveform smoothing as it was described in paragraph
  \ref{gauss_smooth};
  \item Search for the number of pulses and their places in a waveform using
  two user-specified thresholds --- the second derivative threshold $\rm U''$ and
  the deviation threshold $\rm U_{exp}$;
  \item Application of the foregoing algorithms to put the time-marks and to
  measure area of the found pulses;
  \item Correction of pulse area to the contribution from the previous pulses.
\end{itemize}

\begin{figure}
\centering
\includegraphics[width=0.48\textwidth]{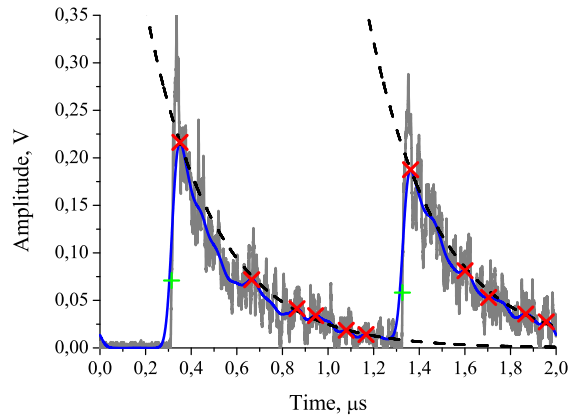}
\caption{An example of processing of an overlapped BGO-pulse. Gray and blue
curves denote the original and smoothed pulses correspondingly. Black dash
curves are theoretical exponents, red and green crosses denote local maximums
of the smoothed pulse and the time-marks.}
\label{figure_intersected_pulse}
\end{figure}

Double numerical differentiation of a waveform was used to find the first
pulse so that the second derivative exceeded the threshold $\rm U''$.
This threshold determines the minimum pulse amplitude (area) that can
be recognized.
So, the tails of the previous pulses can be found, what gives greater
accuracy in the pulse baseline measurement.
On the next step an exponent with the decay time constant $\tau$
from Eq.~\eqref{simple_curve} was drawn from the amplitude of the smoothed
pulse.
A pile-up event was registered when one or some of calculated deviations from
local maximums of the smoothed pulse to the exponent (red crosses in
\figurename\ref{figure_intersected_pulse}) exceeded the value of the threshold
$\rm U_{exp}$. A new exponent was drawn from the amplitude of the found pulse
and so on.
The value of $\rm U_{exp}$ determines the minimum
area of an overlapped pulse, which can be recognized by this algorithm.
Finally, the contribution to the current pulse area from the previous pulses was
calculated as follows:

\begin{equation}
E^* =K\cdot(e^{-\frac{t_2}{\tau}}-e^{-\frac{t_1}{\tau}})
\end{equation}

\noindent where $E^*$ --- the area contribution from a previous pulse,
$K=E_{i-1}/(1-e^{-1})$ --- calibration coefficient, $E_{i-1}$ --- the area
value of the previous pulse measured within the integration time window
$0\ldots300$~ns for the smoothed pulse. The points $t_2$ and $t_1$ are the
boundaries of the integration time window for a current i-th pulse.
If a current pulse was preceded by several pulses, only the area of the last
one was measured, because the previous pulses had already influenced on it.

In the plastic detector pile-up events were unlikely, because pulses
coming from the plastic detector had short time duration ($\sim30$~ns) and a
distance from the detector to a source was 50~cm. 
So only the second derivative threshold $\rm U''$ was used to find the number of
pulses in waveforms and all found maximums of second derivative above the
threshold are considered belong to independent pulses.

\section{Results and discussion}

\subsection{Calibration and energy resolution of the BGO-detectors,
$E_{\gamma}<5$~MeV}

\begin{figure}
\centering
\includegraphics[width=0.48\textwidth]{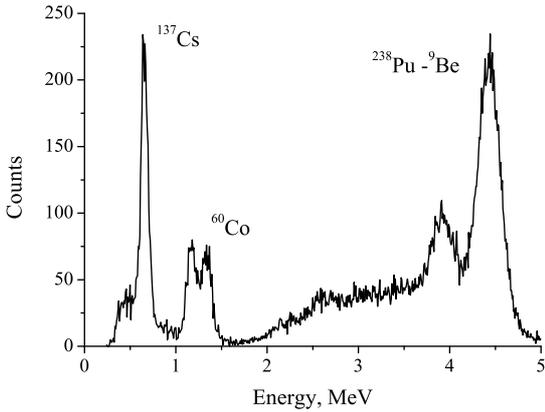}
\caption{The energy spectrum of standard $\gamma$-ray sources of $\rm ^{137}Cs$,
$\rm ^{60}Co$ and $\rm ^{238}Pu-$$\rm ^9Be$ measured by the BGO-detector
coupled with a Tektronix TDS 7704B oscilloscope.}
\label{figure_spectrum_digital}
\end{figure}

\begin{figure}
\centering
\includegraphics[width=0.48\textwidth]{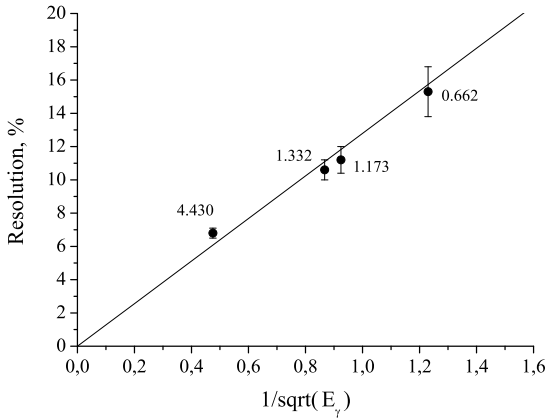}
\caption{Measured energy resolution of the BGO-detector against
$1/\sqrt{E_{\gamma}}$. The solid line shows the straight line fit passing
through the origin.}
\label{figure_sqrt_E}
\end{figure}

The BGO-detectors were calibrated using laboratory standard low energy
$E_{\gamma}<5$~MeV $\gamma$-ray sources: $\rm ^{137}Cs$, $\rm ^{60}Co$ and
$\rm ^{238}Pu-$$\rm ^9Be$. Its spectra are presented in
\figurename\ref{figure_spectrum_digital}. Two peaks in the measured
energy spectrum of a PuBe source were used for calibration --- the total
absorption peak (4.43~MeV $\gamma$-line)  
and the single escape peak (3.92~MeV). The values of the obtained energy
resolution of the BGO-detectors were 15\%, 11\% and 6.8\% for 
a $\rm ^{137}Cs$, a $\rm ^{60}Co$ and a PuBe 
sources, respectively.  The energy resolution was plotted
in \figurename\ref{figure_sqrt_E} as a function of $1/\sqrt{E_{\gamma}}$.
A straight line passing through the origin is fitted to the points.

\begin{figure}
\centering
\includegraphics[width=0.44\textwidth]{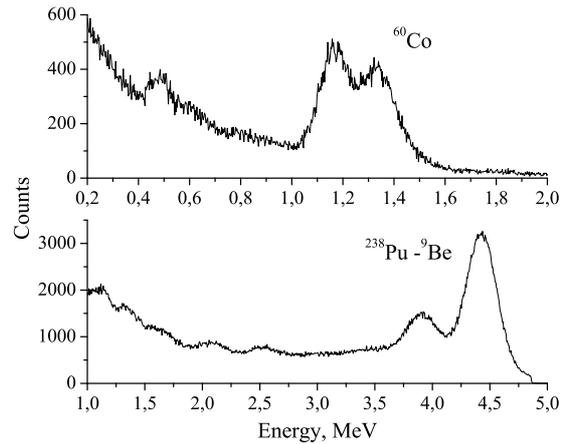}
\caption{The energy spectrum of a $\rm ^{60}Co$ and a $\rm ^{238}Pu-$$\rm ^9Be$
sources measured by the BGO-detector coupled with a 10-bit CAMAC ADC.}
\label{figure_spectrum_analog}
\end{figure}

Also the energy resolution value of the BGO-detectors was measured
using a 10-bit CAMAC ADC. The measured energy spectra of 
a $\rm ^{60}Co$ and a $\rm ^{238}Pu-$$\rm ^9Be$ sources obtained by the
CAMAC apparatus are presented in \figurename\ref{figure_spectrum_analog}. The
energy resolution of the BGO-detectors was 13\% and 7.4\% for a $\rm ^{60}Co$
and a PuBe sources respectively.

In the method based on the application of a fast digital oscilloscope
the pulse area measurement compensates the worse amplitude resolution by means of a high sampling time
discrimination rate. A 10-bit ADC selects the amplitude value by division
to the number of ADC-channels $N_{Amp}=10^{10}$ and a digital oscilloscope with
an 8-bit ADC and $10^4$ sampling points per waveform selects the pulse area by
division on the number of elementary cells $N_{Amp}\times N_T=10^8\times10^4=10^{12}$.
That is the reason why the spectrometer based on a fast
digital oscilloscope with an 8-bit ADC showed a $\sim10\%$ advantage in the
energy resolution over the analog apparatus using a 10-bit ADC with the same detectors and
at the same distance from the detector to a source.
Another advantage of the digital processing method is an opportunity to process
the pulse shapes many times with the use of different numerical algorithms and to develop
 optimal ones.

\subsection{Calibration of the BGO-detectors by cosmic muons}

\begin{figure}
\centering
\includegraphics[width=0.43\textwidth]{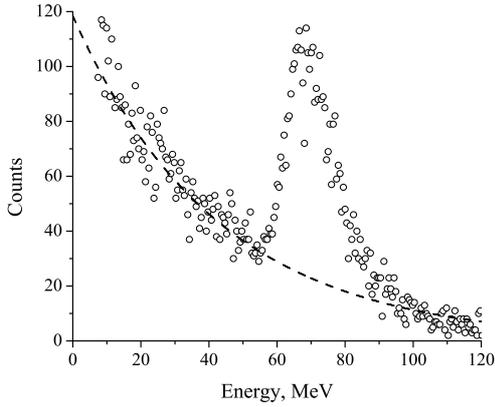}
\caption{The energy spectrum of cosmic rays measured by the BGO-detector. The
peak is corresponded to energy losses of cosmic muons. Exponent fits the falling
part of the spectrum, which is thought not belong to muons.}
\label{figure_spectrum_muon}
\end{figure}

In the energy range $E_{\gamma}>5$~MeV the BGO-detectors were calibrated by
energy losses from cosmic muons in the BGO-crystals. The
BGO-detector responce to muons is appoximately equal to that to
$\gamma$-rays (in Ref.~\cite{Bakken_1994} it was shown that calibration
constants $C_e$ and $C_{\mu}$, defined as the ratio between the energy $E$
released into a crystal and the corresponding measured electric signal $A$, 
are appoximately equal and $(C_e-C_{\mu})/C_{\mu}\approx1\ldots2\%$). 
BGO-crystal was placed vertically to detect muons passing 
through the scintillator along its axis in coincidences
with the plastic detector, which was placed above the BGO-detector.
The measured energy spectrum of cosmic rays is shown in
\figurename\ref{figure_spectrum_muon} where the peak is corresponded to
energy losses from cosmic muons while the exponent tail
is thought belong to cosmic $\gamma$-rays, because an anticoincidence shield was not used. The value of
muon energy losses in the BGO-crystal was calculated 
 using a Monte-Carlo simulation and was equal to $E_{\gamma}\approx68$~MeV.

\begin{figure}
\centering
\includegraphics[width=0.47\textwidth]{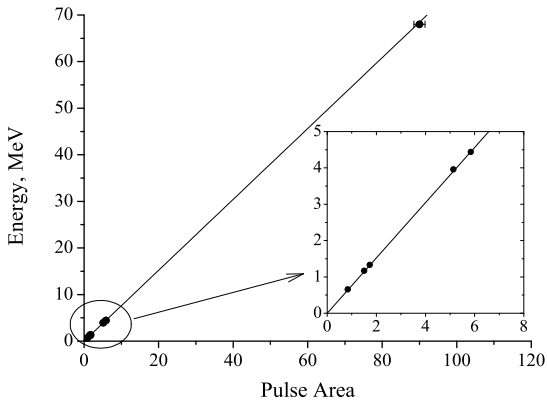}
\caption{The energy calibration curve for the BGO-detector
obtained with standard $\gamma$-ray sources and cosmic muons. Straight line fit
to the low energy points (inset) is extrapolated up to 70 MeV.}
\label{figure_calibration_spectrometer}
\end{figure}

\begin{figure}
\centering
\includegraphics[width=0.47\textwidth]{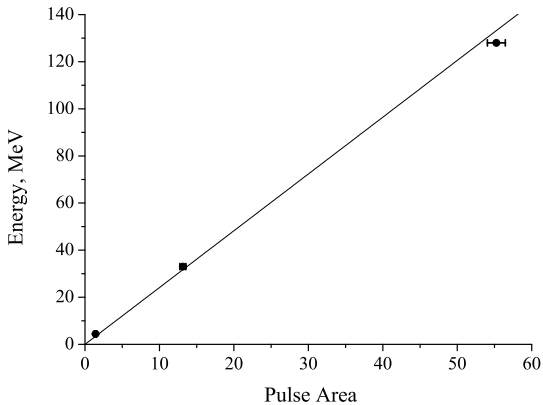}
\caption{The energy calibration curve for the ``long'' BGO-crystal obtained
with a PuBe source and cosmic muons.}
\label{figure_calibration_bgo_long}
\end{figure}

The energy calibration curve for the BGO-detector is presented in
\figurename\ref{figure_calibration_spectrometer}. The fitting of the
experimental points by a straight line was done up to 70 MeV. 
To test the linearity of the calibration curve in the energy range
$E_{\gamma}>70$~MeV a ``long'' $\varnothing\,4\times14.1$~cm BGO-crystal was
used. This crystal was placed vertically and horizontally so that muons passed
different thickness of the BGO crystal along and across its axis.
Monte-Carlo simulations showed that in the case of vertical
oriented position the maximum of the muon peak is corresponded to
$E_{\gamma}\approx128$~MeV of energy loss in the BGO crystal and in the
case of horizontal oriented position to $E_{\gamma}\approx33$~MeV.

The energy calibration curve for the ``long'' BGO-crystal is presented in
\figurename\ref{figure_calibration_bgo_long} where straight line fit passing
through the origin was plotted across the following three points:
$E_{\gamma}=4.43$~MeV from a PuBe source, $E_{\gamma}=33$ and 128~MeV using
cosmic muons at vertical and horizontal oriented positions of the 
detector. The calibration curve was found to be
linear up to $\approx130$~MeV within the measurement errors.

\begin{figure}
\centering
\includegraphics[width=0.47\textwidth]{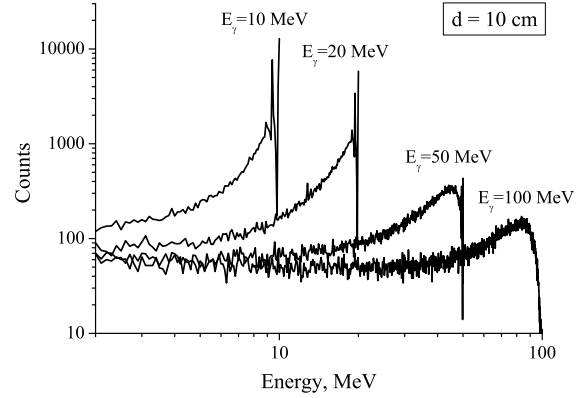}
\caption{The response functions of the BGO-detector to $\gamma$-rays calculated
using Geant4-based Monte-Carlo simulations. A distance from the detector to a
point source was 10 cm.}
\label{figure_geant_10cm}
\end{figure}

\begin{figure}
\centering
\includegraphics[width=0.51\textwidth]{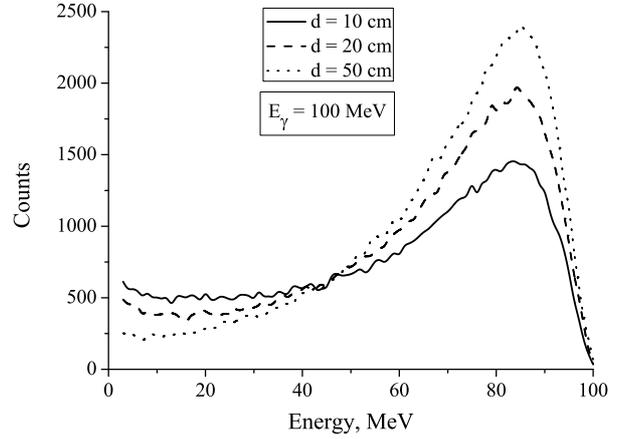}
\caption{The response functions of the BGO-detector to $\gamma$-rays
with energy $E_{\gamma}=100$~MeV calculated using Geant4-based Monte-Carlo
simulations for different distances from the BGO-detector to a point source.}
\label{figure_geant_100MeV}
\end{figure}

\subsection{Response function investigation}

A Monte-Carlo simulation was used to investigate the response
function of the BGO-detectors to $\gamma$-rays.
The calculated responce functions to the $\gamma$-ray energies $E_{\gamma}$ =
10, 20, 50 and 100~MeV are presented in \figurename\ref{figure_geant_10cm}.
A distance between the leading edge of the BGO crystal and a point
$\gamma$-ray source was 10~cm. For each value of $E_{\gamma}$ the same number of
emitted $\gamma$-rays were generated. In \figurename\ref{figure_geant_10cm} it
is shown that the total absorption peak (TAP) decreases with the increasing of
$E_{\gamma}$.
The ratio of the number of events in the TAP to the total number of
detected events was equal to 25.6\%, 9.8\%, 0.6\% and 0.02\% for
$E_{\gamma}$=10, 20, 50 and 100~MeV respectively.
So, we suggest that the energy $E_{\gamma}$ equal to one hundred~MeV will be
an upper limit for this type of $\gamma$-spectrometer.

At second, the dependence of the response function on $\gamma$-rays with fixed
energy $E_{\gamma}=100$~MeV at different distances from the leading edge of
the BGO crystal to a point $\gamma$-ray source was investigated. The calculated
responce functions for three values of distance (10, 20 and 50~cm) are shown in
\figurename\ref{figure_geant_100MeV}. The influence of edge effects (incomplete
$\gamma$-rays absorption due to emission of secondary $\gamma$-rays and
electrons outside the detector) on the shape of the response function is clear.

Also it was shown using a Geant4-based Monte-Carlo simulation that high
energy $\gamma$-rays detection efficiency of a $\varnothing\,7.62\times7.62$~cm BGO
scintillator is approximately equal to the one for a
$\varnothing\,38\times22$~cm $\rm NaI$ scintillator from Ref.~\cite{Luke_1991}.
Therefore, the ratio of the maximum cross-sectional area of the NaI crystal
to the BGO one will be $\sim10$, so the contribution of cosmic ray events will
be $\sim10$ times less for the BGO-crystal.

\subsection{Testing of pile-ups processing }

\begin{figure}
\centering
\includegraphics[width=0.47\textwidth]{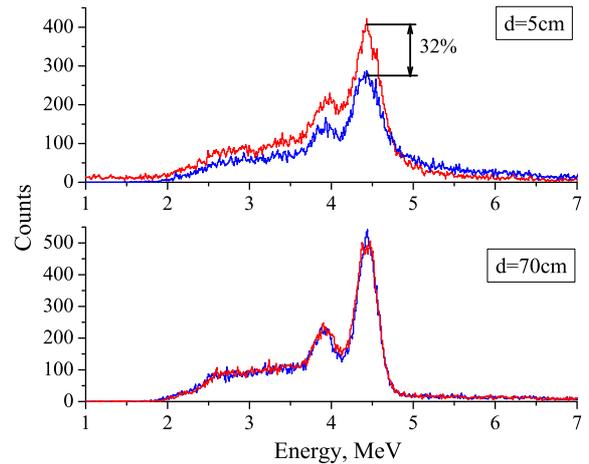}
\caption{The energy spectrum of the $\rm ^{238}Pu-$$\rm ^9Be$ source
obtained by processing of experimental data by two algorithms: the original
one (red curve) and the simplified one (blue curve). Distanses from the
BGO-detector to the source were $d=5$ and 70~cm.}
\label{figure_intersect_prove}
\end{figure}

The algorithm for processing of pile-ups which was mentioned above in paragraph
\ref{algorithm_of_processing_of_pileups} (called the ``original''
algorithm) was tested using one BGO-detector and a PuBe source with an activity of
$4.7\times10^6$~neutron/sec. The PuBe source was placed at eight different
distancies $d$ (from 5 to 70~cm) from the detector. $3\times10^4$ waveforms were collected
for each distance to test efficiency of the ``original'' algorithm. The
obtained experimental data were also processed by the ``simplified'' algorithm which
found only one pulse with maximum amplitude in each waveform. The comparison of
the data processing results for the PuBe energy spectrum is presented in
\figurename\ref{figure_intersect_prove} for two values $d = 5$ and 70~cm.

Both algorithms demonstrated similar results at $d = 70$~cm. 
At $d = 5$~cm the total absorption peak of the $\gamma$-spectrum contained
$32\pm1\%$ greater events in the case of processing by the ``original''
algorithm.
A theoretical estimate of the overlapped pulses portion is the same value
(31\%). For $d=70$~cm a theoretical estimate
is 0.05\% portion of pile-ups that is in agreement with the experimental result
($0.6\pm0.7\%$).
So, the ``original'' algorithm was successfully tested under a counting rate
up to $1.2\times10^5$~waveforms/sec for the BGO-detector (measured
at $d=5$~cm) and the portion of pile-up events more than 30\%.
 
The maximum counting rate of the oscilloscope with the used-specified
setup (the horizontal time window duration 2 $\mu$s, $10^4$ points per waveform)
was measured to be $1.8\times10^5$~waveforms/sec. The measurements were
performed using a digital functional generator Tektronix AFG 3102, which sent a
rectangular pulse train with a 1\% duty cycle to the oscilloscope input.

\section{Conclusion}

A complete digital spectrometer based on a fast digital oscilloscope was
developed and tested together with numerical algorithms for processing the
shapes of pulses coming in coincidences from two BGO- and one plastic
scintillation detectors to distinguish low energy neutrons and $\gamma$-rays by
the time-of-flight method. The time resolution of the
spectrometer was 2.2~ns.
Calibration of the BGO-detectors was performed using standard
$\gamma$-ray sources and cosmic muons. The response function of the
BGO-detectors was investigated using a Geant4-based Monte-Carlo simulation which
showed that there is an upper limit for the use of these detectors related
to the escape effect of total $\gamma$-ray energy absorption in BGO-detectors
at $E_{\gamma}\geq100$~MeV.

Numerical algorithms for digital timing, pulse area measurement and
processing of pile-up events are presented and tested in this work. They
demonstrated a 10\% advantage in the energy resolution of the BGO-detectors over
a 10-bit CAMAC ADC. Also new numerical
algorithm for processing of pile-up events without rejection was successfully
tested for a counting rate    
$1.2\times10^5$~waveforms/sec and the portion of overlapped pulses more than
$30\%$. The maximum counting rate value of $1.8\times10^5$~waveforms/sec was
obtained for the spectrometer.

The developed digital methods of acquisition and processing of the pulse shapes
are universal, they are useful for solving a variety of practical
problems --- for the study of the amplitude-time properties of promising
scintillators \cite{Eremin_2010, Markochev_2013}, rare nuclear
transformations (bremsstrahlung emission accompanying $\alpha$-decay and
fission of heavy nuclei, in heavy ion reactions) and so on.



\section*{References}
\bibliographystyle{elsarticle-num} 
\bibliography{spectrometer}
\biboptions{sort&compress}





\end{document}